\documentclass[twocolumn,prb,aps,amsmath,showpacs]{revtex4}
\usepackage{graphicx}
\usepackage[version=3]{mhchem}
\usepackage{dcolumn}
\usepackage{multirow}
\usepackage{bm}
\usepackage{subfigure}
\usepackage{amssymb}

\begin{document}

\title{Structural stability, vibrational and bonding properties of potassium 1,1$'$-dinitroamino-5,5$'$ bistetrazolate: An emerging green primary explosive}
\author{N. Yedukondalu, and G. Vaitheeswaran$^*$ }
\affiliation{Advanced Centre of Research in High Energy Materials (ACRHEM),
University of Hyderabad, Prof. C. R. Rao Road, Gachibowli, Telangana state, Hyderabad- 500 046, India.}
\date{\today}

\begin{abstract}
Potassium 1,1$'$-dinitroamino-5,5$'$ bistetrazolate (K$_2$DNABT) is a nitrogen rich (50.3 $\%$ by weight, \ce{K2C2N12O4}) green primary explosive with high performance characteristics namely velocity of detonation (D = 8.33 km/s), detonation pressure (P = 31.7 GPa) and fast initiating power to replace existing toxic primaries. In the present work, we report density functional theory (DFT) calculations on structural, equation of state, vibrational spectra, electronic structure and absorption spectra of K$_2$DNABT. We have discussed the influence of weak dispersive interactions on structural and vibrational properties through the DFT-D2 method. We find anisotropic compressibility (b $<$ a $<$ c) from pressure dependent structural properties. The predicted bulk modulus reveals that the material is harder than cyanuric triazide (\ce{C3N12}) and softer than lead azide (\ce{Pb(N3)2}). A complete assignment of all vibrational modes has been made and compared with the available experimental results. The calculated zone centre IR and Raman frequencies show a blue-shift which leads to a hardening of the lattice upon compression. In addition, we have also calculated the electronic structure and absorption spectra using the recently developed Tran Blaha-modified Becke Johnson potential. It is found that K$_2$DNABT is a direct band gap insulator with a band gap of 3.87 eV and the top of the valence band mainly dominated by $2p$-states of oxygen and nitrogen atoms. K$_2$DNABT exhibits mixed ionic (between potassium and tetrazolate ions) and covalent character within tetrazolate molecule. The presence of ionic bonding suggests that the investigated compound is relatively stable and insensitive than covalent primaries. From the calculated absorption spectra, the material is found to decompose under ultra-violet light irradiation.
\end{abstract}

\maketitle
\section{Introduction}
Primary explosives are sensitive towards external stimuli such as impact, heat, friction, electric charge etc.\cite{Huynh1}, showing a very rapid deflagration to detonation transition ($\emph{i.e.}$ reaction front becomes supersonic) by generating a shock wave, which makes the transfer of detonation to a less sensitive secondary explosive.\cite{Klapotke1,Oyler} Consequently, the primary explosives can find applications as initiators (in detonators, primers and blasting caps) for secondary booster charges, main charges or propellants. In addition, primary explosives are vital to any military's portfolio and are essential to construction and mining industries.\cite{Giles} The first primary explosive Mercury (II) Fulminate (MF) adopted by Alfred Nobel and is widely used in commercial blasting caps.\cite{Klapotke1} However, MF is more sensitive, tends to be dead pressed and its base metal Mercury is more expensive. As a result MF finds limited applications in energetic formulations, which calls for a replacement. After an extensive research for about a century, an appropriate replacement for MF were found to be Lead Azide (LA) and Lead Styphnate (LS) which are thermally stable primary explosives. LA is used in detonators due to its high performance (D = 5.92 km/s and P = 33.8 GPa) and high thermal stability (315$^o$C), whereas LS is used in propellant applications because of its low performance. Although LA and LS find tremendous applications in various fields, the presence of heavy metal severely affects the environment and shows a strong influence on any populations that are in close proximity to areas where explosives are set off or fire arms are employed.\cite{Giles} The Pb and Hg based explosives cause damage to the central nervous system, kidneys, heart, brain (human memory) and eyes. This impact is not only limited to soldiers in the battle field, firing ranges ammunition, explosives workers, miners and construction workers are also placed at risk.\cite{Ilyushin} Therefore it is mandatory to search for Hg and Pb free green primary explosives with explosive performance as good as LA and the materials should follow the criterion proposed by Los Alamos national laboratory (LANL).\cite{Huynh1,Ilyushin,Huynh2,Klapotke2} The seven "green" primary criteria are as follows: 1) Insensitivity to moisture and light, 2) Sensitivity to initiation but not too sensitive to handle and transport, 3) Thermally stable at least up to 200 $^o$C, 4) Chemically stable for extended periods, 5) Devoid of toxic metals such as lead, mercury, barium and antimony. 6) Free from perchlorate which may act as a teratogen and has adverse effects on the function of the thyroid gland and 7) Ease and safety of synthesis. Huynh and co-workers\cite{Huynh1, Huynh2} synthesized new coordination complex green primary explosives (5-Nitro tetrazolate N$^2$-ferrate hierarchies) and they replaced the toxic Hg and Pb based primary explosives after about 400 years. Moreover, recently synthesized primary explosives namely copper (I) 5-Nitro tetrazolate (DBX-I)\cite{Fronbarger2} and potassium 5,7-dinitro [2,1,3] benzoxadiazo 1-4-olate 3-oxide (KDNP)\cite{Fronbarger1} also meet the above mentioned LANL criteria for green primaries. Also DBX-1 and KDNP are replacements for LA and LS, respectively. In addition, the KDNP has been qualified by the Department of the Navy as suitable candidate for use as a primary explosive in weapons development while DBX-1 is under evaluation by the department of Navy.\cite{Fronbarger2,Fronbarger1} However, high and low concentration of Copper (using DBX-1) is harmful to living organisms and ecological systems.\cite{Ilyushin} Hence optimal copper should be maintained in the DBX-1 explosive. Very recently a green new primary explosive, potassium 1,1$'$-dinitroamino -5,5$'$ bistetrazolate (K$_2$DNABT) has been synthesized and it exhibits fast detonating characteristics (D = 8.33 km/s and P = 31.7 GPa) and high initiating power as comparable with LA.\cite{Fischer} The results on structure, equation of state and lattice dynamics would lead to fundamental knowledge of the material in future explosive formulations. Therefore, we have computed the pressure dependent structural and vibrational properties as well as electronic and optical properties of the promising green primary explosive. The rest of this article is organized as follows: in section II, we briefly describe the methodology of computation. Results and discussion concerning structural, vibrational properties at ambient as well as at high pressure and electronic structure and optical properties of K$_2$DNABT are presented in section III. Finally section IV summarizes the conclusions of this  study.

\section{Method of computation}
The accuracy and robustness of density functional theory (DFT) in combination with high computing capacity provides remarkable power to computational experiments at the atomistic level. In the present work we have used two distinct approaches to perform \emph{ab-initio} calculations. First one, plane wave pseudo potential approach which is implemented through Cambridge Series of Total Energy Package (CASTEP)\cite{Payne} in Materials Studio. CASTEP is one of the best module in MS which can be used to calculate the physical properties of diverse materials. We have used Vanderbilt\cite{Vanderbilt} ultra-soft (US) and Norm-conserving (NC)\cite{Troullier} pseudo potentials (PPs) for electron-ion interactions from the MS database. The USPPs are used to calculate the structural properties at ambient as well as at high pressure while NCPPs have been used to calculate the pressure dependent zone centre IR and Raman spectra of K$_2$DNABT as they were well suited for lattice dynamical calculations. The local density approximation\cite{Ceperley,Perdew} and generalized gradient approximation (GGA) parameterized by Perdew-Burke-Ernzerhof (PBE)\cite{Burke} were used to treat electron-electron interactions. The Broyden-Fletcher-Goldfarb-Shanno (BFGS) minimization scheme\cite{Almlof} has been used for structural relaxation. The convergence criteria for structural optimization was set to ultra fine quality with a kinetic energy cut-offs of 600 eV for USPPs, 950 eV for NCPPs and 5 $\times$ 4 $\times$ 3 k-mesh according to the Monkhorst-Pack grid scheme.\cite{Monkhorst} The self-consistent energy convergence less than 5.0$\times$10$^{-6}$ eV/atom and maximal force between atoms was set to 0.01 eV/$\textup{\AA}$. The maximum displacement and stress were set to be 5.0$\times$10$^{-4}\textup{\AA}$ and 0.02 GPa, respectively.

Semi-empirical dispersion correction method (DFT-D2) by Grimme\cite{Grimme} was applied to treat weak dispersive interactions. Therefore, the total energy after inclusion of dispersion correction is given by
\begin{equation}
E_{DFT+D} = E_{DFT} + E_{disp}
\end{equation}
Where E$_{DFT}$ is the self-consistent Kohn-Sham energy, E$_{disp}$ is the empirical dispersion correction energy which is given by
\begin{equation}
E_{disp} = -S_6\sum\limits_{i \textless j}\frac{C_{ij}}{R_{ij}^6}f_{damp}(R_{ij})
\end{equation}
Where S$_6$ is the global scaling factor that only depends on the density functional used. C$_{ij}$ denotes the dispersion coefficient for the pair of i$^{th}$ and j$^{th}$ atoms that depends only on the chemical species, and R$_{ij}$ is an inter atomic distance. f$_{damp}$ = $\frac{1}{1+e^{-d(R_{ij}/R_0-1)}}$ is a damping function which is necessary to avoid divergence for small values of R$_{ij}$ and R$_0$ is the sum of atomic van der Waals (vdW) radii. Moreover, it is well known from the literature that DFT-D2 method works well for the solid energetic molecular crystals.\cite{konda} Therefore, we systematically investigated the effect of vdW interactions on structural and vibrational properties of the layered and molecular crystal K$_2$DNABT.

Second one, full potential linearized augmented plane wave (FP-LAPW) method which is implemented in WIEN2K package.\cite{blaha} It is well known fact that The standard DFT functionals severely underestimate the band gap by 30-40$\%$ for semiconductors and insulators. In contrast to LDA/GGA functionals, recently developed Tran-Blaha modified Becke Johnson (TB-mBJ)\cite{peter} potential shows remarkable success in predicting the energy band gaps for diverse materials\cite{singh,camargo,dixit,jiang} and competing with the computationally expensive methods such as GW approximation, hybrid functionals. Therefore, in the present work, TB-mBJ potential has been used to get reliable band gap thereby calculation of electronic structure and optical properties for the investigated compound. To achieve the required convergence of energy eigenvalues, the wave functions in the interstitial region were expanded using plane waves with a cut-off K$_{max}$ = 7/RMT while the charge density was Fourier expanded up to G$_{max}$ = 14, where Radius of Muffin Tin (RMT) is the smallest atomic sphere radius and K$_{max}$ denotes the magnitude of the largest K vector in plane wave expansion. The RMT radii are assumed to be 1.2, 1.12, 1.14 and 1.16 Bohrs for K, C, N and O, respectively. The wave functions inside the spheres are expanded up to $l_{max}$ = 10. Self-consistency of total energy is obtained by using 5 $\times$ 4 $\times$ 3 k-mesh in the Irreducible Brillouin Zone (IBZ). The frequency-dependent optical properties have been calculated using a denser k-mesh of 10 $\times$ 8 $\times$ 6 in the IBZ.

\section{Results and discussion}
\subsection{Crystal structure and equation of state}
K$_2$DNABT crystallizes in the triclinic P$\bar{1}$ symmetry with unit cell parameters a = 5.0963 $\AA$, b = 6.8248 $\AA$, c = 8.44271 $\AA$, $\alpha$ = 67.56$^o$, $\beta$ = 86.15$^o$, $\gamma$ = 71.02$^o$ and Z = 1 at 100 K.\cite{Fischer} The unit cell of experimental crystal structure is shown in Figure 1(a), in which the two potassium atoms are ionically bonded to two tetrazolate anions through a nitrogen atom, the K2DNABT molecule are arranged in a puckered layered manner in xy-plane as shown in Figure 1(b), and a single tetrazolate molecule is shown in Figure 1(c). By considering the experimental structure as an input, we first optimized the lattice geometry by fully relaxing atomic coordinates and lattice constants without treating weak dispersive interactions. The obtained lattice constants using standard DFT functionals are found to differ by $\sim$ 1-3.5$\%$ and $\sim$ 1.2-3.8$\%$ within LDA and PBE-GGA, respectively (see Table I). The predicted equilibrium volume is underestimated by 8.8$\%$ within LDA conversely the same is overestimated by 10.8$\%$ within PBE-GGA. This clearly shows the poor description of dispersive interactions in this material using the standard DFT functionals.
\par Extensive studies have been reported in the literature describing the inability of LDA/GGA functionals in treating these weak dispersive interactions in layered and molecular crystalline solids.[Ref.\onlinecite{Sorescu} and Refs there in] In order to capture these weak dispersive forces, semi-empirical dispersion correction methods have been developed and implemented through the standard DFT description. The goal of the dispersion correction methods is to enable DFT to have predictive power for large assemblies of molecules. There are two methods of such corrections; the first one is pairwise additive correction and the second one is non-local correlation and recently both of these methods have shown remarkable success.\cite{Cho} In the present study, we have used DFT-D2 method which is based on pairwise additive correction. The calculated lattice constants deviate by 0.1-0.6$\%$ and the corresponding volume is predicted to differ by +1.2$\%$. This is small in magnitude when compared to standard DFT errors and the results are in good agreement with the experiments\cite{Fischer} as depicted in Table \ref{tab:struct}. There is a significant improvement with DFT-D2 method over standard DFT calculations without dispersion corrections. Also the obtained intra molecular C-C, C-N (C=N), N-N (N=N) and N-O interactions and torsion angles are now closely comparable with the experimental results\cite{Fischer} as displayed in Table \ref{tab:lengths}. Further, the optimized structure at ambient pressure is used to perform high pressure structural and vibrational properties of K$_2$DNABT.

\par High pressure studies on structural properties of ionic layered compounds give a good insight on how pressure affects the weak and strong bonds in condensed matter. We have systematically investigated the structural properties of K$_2$DNABT at zero Kelvin temperature under hydrostatic pressure up to 10 GPa in steps of 1 GPa. As illustrated in Figure 2(a), lattice constants a, c decrease monotonically with increasing pressure while the b lattice constant decreases up to 4 GPa and further it shows large compression in the pressure range of 4-10 GPa. Comparison of the calculated lattice constants at 0 and 10 GPa shows the reduction in lattice constants a, b, and c by 0.33, 1.01, and 0.44 $\AA$, respectively. It can be clearly noticed that the reduction in b-axis is much larger than that of a $\&$ c-axes in the studied pressure range. More precisely, we plotted the normalized lattice constants as a function of pressure as displayed in Figure 2(b); apparently shows the lattice constants a, b and c shrink with different compressibility 93.52 $\%$, 85.17 $\%$, and 94.78 $\%$, respectively. Apart from this, we have also calculated the first order pressure coefficients from the following relation $\gamma($X$)$ = $\frac{1}{X}\frac{dX}{dP}$, ($X$ = a, b, c) by fitting lattice constant and pressure data to fifth order polynomial. The first order pressure coefficients (in GPa$^{-1}$) are found to be 10.3 $\times$ 10$^{-3}$, 20.4 $\times$ 10$^{-3}$, and 13.7 $\times$ 10$^{-3}$ for lattice parameters a, b, and c, respectively. The compressibility of a lattice constant can be correlated with the strength of intermolecular interactions observed along that axis. One would expect much lower compressibility for axes that involve strong intermolecular interactions and the corresponding longitudinal elastic stiffness constant is found to be higher along that crystallographic axes.\cite{Haycraft} Further this can be used to correlate with the detonation sensitivity of an explosive material along particular crystallographic direction. From the calculated compressibility and pressure coefficients, the response of lattice constants to the applied pressure varies as follows b$\textless$a$\textless$c, which clearly indicates that b- and and c- axes are most and least compressible respectively. This implies that K$_2$DNABT is found to be most sensitive to detonate along crystallographic b-axis and least sensitive along the c-axis. The calculated lattice angles and their normalized values are plotted as a function of pressure as shown in Figure 2(c$\&$d). Variation in the lattice angles are found to be less pronounced with pressures up to 3 GPa for $\alpha$ and $\gamma$, and up to 10 GPa for $\beta$; but above 3 GPa, $\alpha$ and $\gamma$ angles show an increment with application of pressure up to 10 GPa. Also a discontinuity is observed in the lattice constant 'b' and lattice angles $\alpha$, $\gamma$ in the pressure range of 4-6 GPa. Moreover, the equations of state are important in describing the properties of materials. Therefore, we have calculated the volume as a function of pressure and found that it decreases monotonically under studied pressure range as shown in Figure 2(e). We could not see any discontinuity in the P-V curve and the ambient volume is compressed to 21.7 $\%$ at 10 GPa (see Figure 2(f)). To predict the bulk modulus (B$_0$) and its pressure derivative (B$_0'$), the calculated compression data were fitted to the third-order Birch-Murnaghan equation of state.\cite{Birch} The bulk modulus and its pressure derivative are found to be B$_0$ = 22.4 GPa and B$_0'$ = 4.7, respectively. The obtained B$_0$ value reveals that K$_2$DNABT is harder than the very sensitive\cite{Agrawal} organic primary explosive \ce{C3N12} (12.6 GPa)\cite{kondaiah} and softer than inorganic primary explosives AgN$_3$ (39 GPa)\cite{hou}, Pb(N$_3$)$_2$ (26 GPa\cite{millar} and 41 GPa\cite{weir}) and other nitrogen rich energetic salts such as 27.2 GPa for KN$_3$\cite{Ramesh} and 26.34 for NH$_4$N$_3$\cite{Kondal}. We also investigated the pressure dependence of intra molecular bond lengths and plotted the normalized bond lengths as shown in Figure 3(a $\&$ b). All these bonds are found to stiffen $i.e.$ decrease monotonically with progression of pressure, among them C-C bond is the most compressible under the studied pressure range. In addition, we also observe a discontinuity in N5-N1 and N6-N5 bonds in the pressure range of 4-6 GPa as seen previously for the lattice constants of unit cell, which might suggest a structural distortion/transition in the K$_2$DNABT crystal in this pressure range. Overall, we observe $\sim$ 1 $\%$ shortening in the bond lengths when compared at 0 and 10 GPa pressures. Also the calculated torsion angles N1-C1-C2-N4 $\&$ O1-N6-N5-N1 are increased by $\sim$ 1.8 times whereas C1-N1-N5-N6 angle reduced to 0.8 times of the values at 0 GPa as shown in Figure 3(c).

\subsection{Zone centre IR and Raman spectra at ambient and under pressure}
In addition to the determination of crystal structure of K$_2$DNABT, Fischer et al\cite{Fischer} also measured IR and Raman frequencies for this material but they have not assigned any of the vibrational modes. Due to complex crystal structure of K$_2$DNABT it becomes a tedious task for experimentalists to assign all the vibrational modes. So we turn our attention to analyse the complete zone centre vibrational spectrum of K$_2$DNABT. The zone centre phonon calculations were carried out using linear response approach within density functional perturbation theory (DFPT). The linear response approach allows one to obtain the effective charges and dielectric tensors directly without any super cell. Single crystal X-ray diffraction study reveals that K$_2$DNABT crystallizes in the triclinic P$\bar{1}$ symmetry with one formula unit which includes 20 atoms per unit cell resulting in 60 (3 acoustic + 57 optical) vibrational modes. Using the method of factor group analysis, we determine the distribution of the zone-centre vibrational modes in terms of the representation of C$_i$ point group as $\Gamma_{total}$ = 30A$_u$  $\oplus$ 30A$_g$ ($\Gamma_{acoustic}$ = 3A$_u$;  $\Gamma_{optic}$ = 27A$_u$  $\oplus$ 30A$_g$). Since P$\bar{1}$ space group has inversion symmetry, IR and Raman modes do not mix; thus the A$_g$ modes are Raman active whereas A$_u$ modes are IR active, in which $g$ and $u$ represent symmetric and antisymmetric modes with respect to centre of inversion, respectively. The calculated zone centre IR and Raman active vibrational modes and their complete vibrational assignment are given in Table \ref{tab:vib}. The vibrational spectrum of K$_2$DNABT can be divided into four regions. 1) The low frequency modes between 60-200 cm$^{-1}$ which arise due to both potassium cation and tetrazolate anion. 2) The bands from 200-500 cm$^{-1}$ are mainly due to rotational and translational motions of tetrazole ring and various branches of anion N-N-N, N=N-N, and NO$_2$ group within and outside the tetrazole ring. 3) Phonon branches in the energy range 500-1090 cm$^{-1}$ mainly due to twisting, bending and scissoring motion of C(N)=N-N, N-NO$_2$, C-C and C-N bands. 4) The bands from 1095-1560 cm$^{-1}$ mainly due to symmetric and asymmetric stretching of NO$_2$, N-N=N, C-N bands.
\par We also extended our calculations beyond the experiments to investigate IR and Raman spectra under pressure up to 10 GPa. The calculated IR spectra at ambient as well as at high pressure are displayed in Figure 4. As illustrated in Figure 4, the frequency and intensity of the IR modes increase with increasing pressure, which shows the hardening of the crystal under hydrostatic compression. However it is clearly seen from Figure 4a, as pressure increases the intensity of two lattice modes with vibrational frequency 116.0 and 131.1 cm$^{-1}$ decrease and the reduction in their intensities are about 58 $\%$ and 84 $\%$ respectively when compared to ambient and highest pressure carried out in the present study. The experimental\cite{Fischer} ambient IR data explores that there are strong absorption peaks at 1300 and 1440 cm$^{-1}$. The corresponding IR peaks from our present calculations (see Figure 4) are found to be at 1239.3 and 1393.6 cm$^{-1}$, which are in close comparison (differ by $\sim$ 5$\%$) with the experimental data.\cite{Fischer} The two absorption bands 1239.3 and 1393.6 cm$^{-1}$ originated from NO$_2$ symmetric, C-N asymmetric and NO$_2$ asymmetric stretching, respectively. 1064 nm (Nd:YAG) laser has been used as an incident light to measure Raman data\cite{Fischer} for K$_2$DNABT single crystals. To compare with the experimentally measured Raman data, we have also calculated the Raman spectra at 298 K using the same laser source as an incident light. Figure 5(a) reveals that the lattice modes are dominant below 300 cm$^{-1}$ and we observed only one strong peak at 1554.7 cm$^{-1}$ in the high frequency region (see Figure 5(c $\&$ d)) with the 1064 nm laser, which is in good agreement with the experimental value of 1610 cm$^{-1}$. However, the Raman bands correspond to twisting, scissoring and bending modes in the frequency range 900-1450 cm$^{-1}$ are displaying very low intensity which complicates the Raman measurements in determining the bending modes of the material with the 1064 nm laser (see Figure 1 of supplementary material).\cite{support} Hence we have also calculated the Raman spectra using standard 514.5 nm Ar laser and the calculated Raman bands corresponding to bending modes are as shown in Figure 5(b). Using 514.5 nm laser, the internal modes ($\textgreater$ 900 cm$^{-1}$) are found to be more intense (see Figure 5(d)) while the lattice modes ($\textless$ 300 cm$^{-1}$) are found to have more intensity (see Figure 5(a)) using 1064 nm laser. This information could motivates future Raman spectroscopic measurements on this material at ambient and at elevated pressures. Furthermore, effect of hydrostatic compression on Raman spectra is investigated using both of the sources. Our calculations show that the intensity of the lattice modes decreases with pressure and a significant reduction is observed for the lattice mode 76.2 cm$^{-1}$ whereas the frequency as well as intensity of the internal modes grow with application of hydrostatic pressure. However, all these modes keep their identity in the IR and/or Raman spectra up to 10 GPa. The few simulated snapshots of IR and Raman eigenvectors are depicted in Figure 6. Overall we predict that the IR and Raman modes exhibit pressure induced blue shift which is due to the inter and intra molecular interactions which stiffen upon compression. This would suggest the dynamical stability of K$_2$DNABT under the studied pressure range. Our results could be considered as a reference for future experimental or theoretical studies of the green primary explosive at ambient as well as at high pressure.

\subsection{Electronic structure, chemical bonding and absorption spectra}
Electronic structure and optical properties of K$_2$DNABT crystal have been investigated using FP-LAPW method by optimizing the fractional co-ordinates at the experimental lattice constants within PBE-GGA as displayed in Table 1 of the supplementary material.\cite{support} The obtained fractional co-ordinates are in good agreement with the experimental data.\cite{Fischer} The study of microscopic properties like electronic structure and chemical bonding play a vital role in understanding the macroscopic energetic behavior and stability of explosive materials. Band gap is a fundamental parameter of any semiconducting/insulating material. Zhang and co-workers explained the correlation between bond dissociation energy, band gap and impact sensitivity for nitro-aromatic compounds.\cite{Zhang} Following the study, Zhu et al\cite{zhu1} also made a correlation between band gap and impact sensitivity of energetic materials (metal azides and styphanates, pure and K-doped CuN$_3$, polymorphs of HMX, CL-20 and TATB) based on the Principle of Easiest Transition (PET), which states that with a smaller band gap it is easier to transfer the electron from valence band to conduction band. The calculated band gap for K$_2$DNABT crystal is found to be 2.87 eV within PBE-GGA, which is 0.45 eV less that of the band gap value 2.42 eV\cite{zhu2} of $\alpha$-lead azide (LA). According to PET criteria, K$_2$DNABT should be less sensitive than LA but the experimental measurements disclose that K$_2$DNABT is more sensitive than LA. This clearly indicates the band gap and impact sensitivity criteria will be helpful to make correlation within a group of materials and/or between the polymorphs. In order to make this correlation, one has to predict accurate band gaps for the energetic materials. Therefore in the present study we have used TB-mBJ potential to calculate the electronic band gap of the material. The calculated band gap using TB-mBJ potential is 3.57 eV which is improved by 0.7 eV when compared to PBE-GGA value of 2.87 eV and the predicted band gap serves as a reference for future experiments. The TB-mBJ band structure is plotted along high symmetry directions $\Gamma$ (0.0, 0.0, 0.0) $\rightarrow$ F (0.0, 0.5, 0.0) $\rightarrow$ Q (0.0, 0.5, 0.5) $\rightarrow$ Z (0.0, 0.0, 0.5) $\rightarrow$ B (0.5, 0.0, 0.0) $\rightarrow$ $\Gamma$ (0.0, 0.0, 0.0) as shown in Figure 7(a); K$_2$DNABT is a direct band gap insulator along the B direction of the Brillouin zone. In order to understand the contribution of each atom to the electronic band structure, we have plotted the partial density of states (PDOS) for each inequivalent atom of the unit cell. As illustrated in Figure 7(b) the states near the Fermi level are mainly dominated by more electronegative oxygen (O1 $\&$ O2) and nitrogen (N5) atoms. The $2p$-states in the energy range -6 to -8 eV are derived from nitrogen (N6) and oxygen (O1 $\&$ O2) atoms and they have strong hybridisation between N6 and O1(O2) which shows the covalent nature of NO$_2$ group within the unit cell. It is also seen from the PDOS and charge density map (see Figures 7(b $\&$ c)), there is a charge sharing between C-N1, C-N4, N1-N2, N2-N3 $\&$ N3-N4 atoms which implies a covalent bonding between these atoms within the tetrazolate anion. However, there is no charge sharing between potassium cation and tetrazolate anion which clearly indicates that the bonding in this material is strongly ionic. Overall we observed mixed bonding nature in K$_2$DNABT crystal; strong ionic bonding between potassium and tetrazolate ions and also covalent bonding within the tetrazolate anion. X-ray electron spectroscopic study\cite{Colton} on inorganic metal azides reveals that Alkali metal azides (AMAs) possess more ionic character than Heavy Metal azides (HMAs), implying that AMAs are relatively stable than HMAs. On the similar path the presence of ionic bonding in K$_2$DNABT crystal makes it relatively stable and insensitive when compared to covalently bonded primary explosives (cynaruic triazide\cite{Agrawal}). The calculated absorption spectra of the K$_2$DNABT crystal along three crystallographic directions using the TB-mBJ potential is shown in Figure 8. The computed absorption spectra shows that the strong optical anisotropy and absorption starts at the band gap value. As the optical edge starts at 3.57 eV, we confirm that K$_2$DNABT crystal is expected to undergo decomposition by the irradiation of Ultra-Violet (UV) light with a wavelength of 347.3 nm. The excitonic peak may lie below the calculated absorption edge (347.3 nm) and these excitonic effects can be obtained by solving the Bethe-Salpeter equation using quasi particle energies which is a future direction emerging for this study.

\section{Conclusions}
In conclusion, we have investigated the pressure dependent structure, the equation of state and vibrational spectra of the emerging green primary explosive potassium 1,1$'$-dinitroamino -5,5$'$ bistetrazolate. The calculated ground state properties using DFT-D2 are in good agreement with the experiments in contrast to standard LDA/GGA functionals. The anisotropic compressibility (b $<$ a $<$ c) of K$_2$DNABT crystal implies that the material is found to be most sensitive to detonation along crystallographic b-axis and least sensitive along the c-axis. The obtained equilibrium bulk modulus reveals that K$_2$DNABT is softer than toxic lead azide and harder than the most sensitive cyanuric triazide. A complete assignment of all vibrational modes has been made according to their molecular vibrations and compared with the available experimental data. The computed IR and Raman spectra show hardening of the lattice under compression. In addition, we also calculated electronic structure and bonding properties using recently developed TB-mBJ potential and it is found that the compound is a direct band gap insulator with band gap value 3.87 eV. K$_2$DNABT possesses mixed bonding nature with strong ionic bonding between potassium cation and tetrazolate anion and also covalent bonding within the tetrazolate anion. The existence of ionic bonding may suggest that the material is stabler than covalent primary explosives. The computed absorption spectra indicate that the present investigated compound is sensitive to UV light. The present study could be helpful when this material finds applications in energetic formulations as a green primary explosive.

\section{Acknowledgments}
Authors would like to thank Defence Research and Development Organisation (DRDO) through ACRHEM for the financial support under grant No. DRDO/02/0201/2011/00060:ACREHM-PHASE-II, and the CMSD, University of Hyderabad, for providing computational facilities. NYK acknowledges Prof. C. S. Sunandana, School of Physics, University of Hyderabad, Dr. V. Kanchana, Department of Physics, Indian Institute of Technology Hyderabad for critical reading of the manuscript. We would like to thank the anonymous reviewers for the useful comments and suggestions. \\


{\pagestyle{empty}

\begin{table*}[h]
\caption{Calculated ground state lattice constants (a, b, c, $\alpha$, $\beta$, $\gamma$), volume (V), and density ($\rho$) of K$_2$DNABT using standard DFT functionals (LDA, PBE-GGA) and dispersion corrected (DFT-D2) method. Experimental data have been taken from Ref.\cite{Fischer} and the relative errors were given in parenthesis with respect to experimental data. ‘-’ and ‘+’ represent under and overestimation of calculated values when compared to the experiments.}
\begin{ruledtabular}
\begin{tabular}{ccccc}
 Parameter          &       CA-PZ        &         PBE       &        DFT-D2     &  Expt.\cite{Fischer}  \\ \hline
a (\AA)             &  4.995 (-2.0$\%$)  &  5.158 (+1.2$\%$) &  5.127 (+0.6$\%$) &  5.0963 \\
b (\AA)             &  6.583 (-3.5$\%$)  &  7.044 (+3.2$\%$) &  6.832 (+0.1$\%$) &  6.8248 \\
c (\AA)             &  8.224 (-2.4$\%$)  &  8.746 (+3.8$\%$) &  8.474 (+0.5$\%$) &  8.4271 \\
$\alpha$ ($^o$)     &  66.60 (-1.4$\%$)  &  69.46 (+2.8$\%$) &  67.29 (+0.4$\%$) &  67.56  \\
$\beta$ ($^o$)      &  85.35 (-0.9$\%$)  &  87.31 (+1.3$\%$) &  85.92 (+0.3$\%$) &  86.15  \\
$\gamma$ ($^o$)     &  70.15 (-1.2$\%$)  &  72.60 (+2.2$\%$) &  71.18 (+0.2$\%$) &  71.02  \\
V ($\AA^3$)         &  233.0 (-8.8$\%$)  &  283.4 (+10.8$\%$)&  258.7 (+1.2$\%$) &  255.65 \\
$\rho$ (gr/cc)      &  2.382 (+9.7$\%$)  &  1.959 (-9.8$\%$) &  2.146 (-1.2$\%$) &  2.172  \\
\end{tabular}
\end{ruledtabular}
\label{tab:struct}
\end{table*}

\begin{table*}[h]
\caption{Calculated  bond lengths (in, $\textup{\AA}$) and torsion angles (in, $^{\circ}$) of K$_2$DNABT using standard DFT functionals (LDA, PBE-GGA) and dispersion corrected (DFT-D2) method. Experimental data have been taken from Ref.\cite{Fischer}}
\begin{ruledtabular}
\begin{tabular}{ccccc}
Parameter      &   LDA    &   PBE  & DFT-D2 &  Experiment\cite{Fischer}  \\   \hline
Bond lengths  \\
C1-C2          &  1.420   &  1.438 & 1.436  & 1.453 \\
C1-N1          &  1.352   &  1.365 & 1.364  & 1.352 \\
C1-N4          &  1.327   &  1.339 & 1.338  & 1.322 \\
O1-N6          &  1.256   &  1.269 & 1.267  & 1.248 \\
O2-N6          &  1.264   &  1.278 & 1.276  & 1.261 \\
N1-N2          &  1.352   &  1.368 & 1.366  & 1.355 \\
N2-N3          &  1.317   &  1.327 & 1.327  & 1.298 \\
N3-N4          &  1.349   &  1.365 & 1.364  & 1.361 \\
N5-N1          &  1.367   &  1.392 & 1.389  & 1.394 \\
N6-N5          &  1.340   &  1.359 & 1.359  & 1.332 \\
Torsion angles  \\
C1-N1-N5-N6    &  70.7    &  82.6  &  72.8  & 75.3 \\
N1-C1-C2-N4    &  -2.2    &  -1.6  &  -1.6  & -1.5 \\
O1-N6-N5-N1    &   3.7    &  0.05  &  2.7   & 1.6  \\
\end{tabular}
\end{ruledtabular}
\label{tab:lengths}
\end{table*}

\begin{table*}[h]
\caption{Calculated vibrational spectrum and their assignment of K$_2$DNABT at the DFT-D2 equilibrium volume.}
\begin{ruledtabular}
\resizebox{0.99\textwidth}{!}{
\begin{tabular}{cccccc}
        Frequency      &   IrrRep    &    Assignment                &  Frequency    &   IrrRep    &   Assignment    \\ \hline
           60.7        &   A$_g$     &   lattice modes              &   635.8       &   A$_u$     &  C=N-N bend   \\
           66.0        &   A$_u$     &                              &   664.9       &   A$_g$     &  N=N-N bend  \\
           76.2        &   A$_g$     &                              &   682.4       &   A$_u$     &  C=N-N bend  \\
           94.0        &   A$_u$     &                              &   704.8       &   A$_u$     &  NO2 bend \\
           100.7       &   A$_g$     &                              &   705.8       &   A$_g$     &  NO2 bend  \\
           107.7       &   A$_u$     &                              &   724.8       &   A$_g$     &  N-C=N, NO2 bend \\
           109.1       &   A$_g$     &                              &   726.3       &   A$_u$     &  NO2 bend \\
           116.0       &   A$_u$     &                              &   733.5       &   A$_g$     &  C-C, N-O bend \\
           118.6       &   A$_g$     &                              &   827.9       &   A$_u$     &  N-N-O twist \\
           129.6       &   A$_g$     &                              &   842.8       &   A$_g$     &  N-N-O twist  \\
           131.1       &   A$_u$     &                              &   957.6       &   A$_u$     &  N=C-C bend, N-NO2 str  \\
           140.8       &   A$_g$     &                              &   958.2       &   A$_g$     &  N=N-N scissor  \\
           159.6       &   A$_u$     &                              &   966.2       &   A$_u$     &  N-NO2 sym str, N=N-N scissor  \\
           160.6       &   A$_g$     &                              &   966.3       &   A$_g$     &  N-NO2 sym str, N=N-N scissor \\
           163.5       &   A$_u$     &                              &   996.4       &   A$_u$     &  C=N-N asym str, N=N Scissor \\
           179.1       &  A$_u$      & Rot. lattice                 &   1044.7      &  A$_g$   & N=N-N asym str, N-N scissor  \\
           179.7       &  A$_g$      & Rot. N-O                     &   1078.2      &  A$_u$   & N=N-N wagg       \\
           202.2       &  A$_u$      & Rot. lattice                 &   1082.4      &  A$_g$   & N=N-N wagg      \\
           277.2       &  A$_g$      & Rot. N-N-N                   &   1098.8      &  A$_u$   & N=N-N, C-N-N asym str \\
           287.3       &  A$_g$      & Rot. N-N=N                   &   1192.2      &  A$_g$   & C-N-N asym str   \\
           355.9       &  A$_u$      & Trans. C-C=N-N, N-N-N-O bend &  1197.1       &  A$_u$   & N-N=N asym str  \\
           368.9       &  A$_g$      & Rot. ring, N-O               &  1217.1       &  A$_g$   & N-N=N asym str  \\	
           431.7       &  A$_g$      & Rot. C-C, N-O, Trans. C-N    &  1239.3       &  A$_u$   & NO2 asym str, C-N asym str \\
           435.2       &  A$_u$      & Rot. ring, N-N-O             &  1260.9       &  A$_g$   & NO2 sym str  \\
           463.0       &  A$_u$      & Trans. ring,  Rot. N-O       &  1301.0       &  A$_u$   & N=C-N asym str  \\
           495.2       &  A$_g$      & Trans. ring, Rot. N-O        &  1353.5       &  A$_u$   & C-N sym str  \\
           579.6       &  A$_g$      & Rot. ring, C-C twist         &  1378.3       &  A$_g$   & C-N, N-O asym str \\
                       &             &                              &  1393.6       &  A$_u$   & NO2 asym str \\
                       &             &                              &  1393.9       &  A$_g$   & C-N, N-O asym str \\
                       &             &                              &  1554.7       &  A$_g$   & C-C=N(-N) asym str\\	
\end{tabular}}
\end{ruledtabular}
\label{tab:vib}
\end{table*}

}

\clearpage
\begin{figure*}[h]
\centering
\includegraphics[height = 4.5in, width=6.0in]{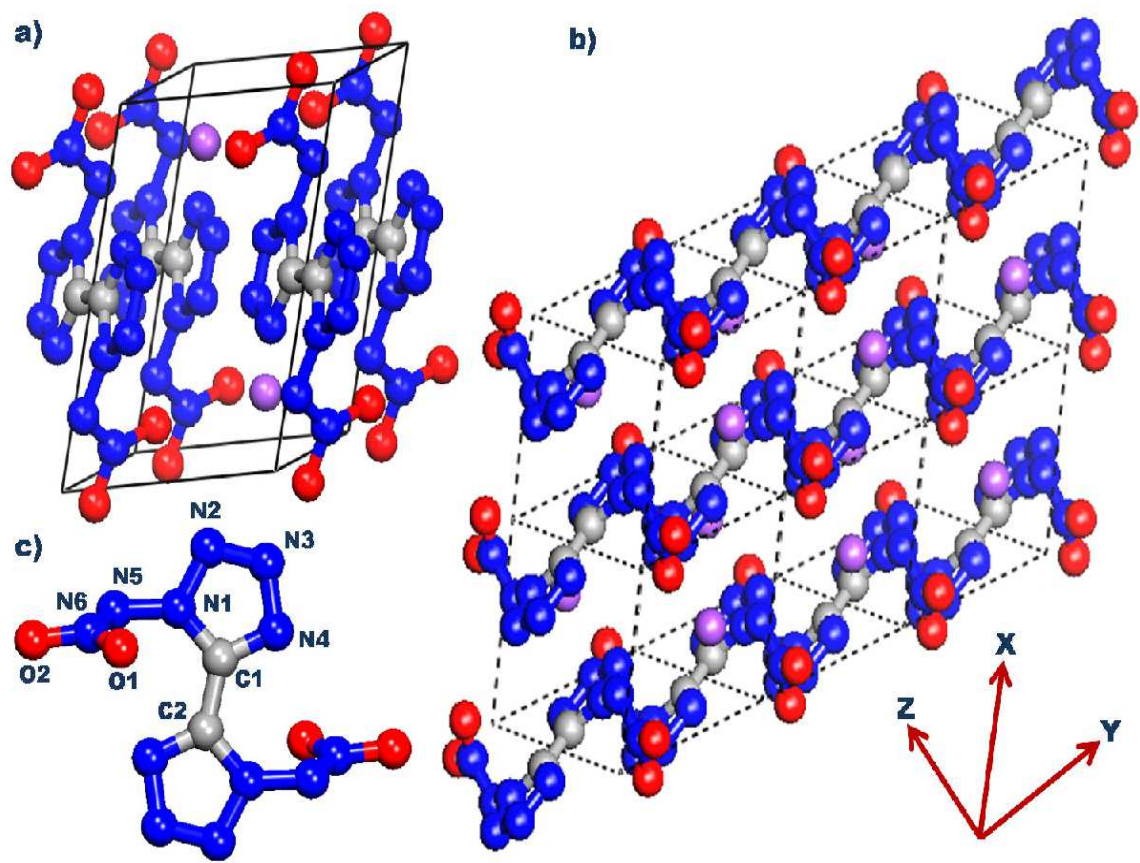}
\caption{(Colour online) (a) Triclinic (P$\bar{1}$) unit cell, (b) Puckered layered structure viewed along xy-plane, (c) Single molecular geometry of tetrazolate anion of the K2DNABT crystal. Ash, blue, red, and violet colour balls represent carbon, nitrogen, oxygen and potassium atoms, respectively.}
\end{figure*}

\begin{figure*}[h]
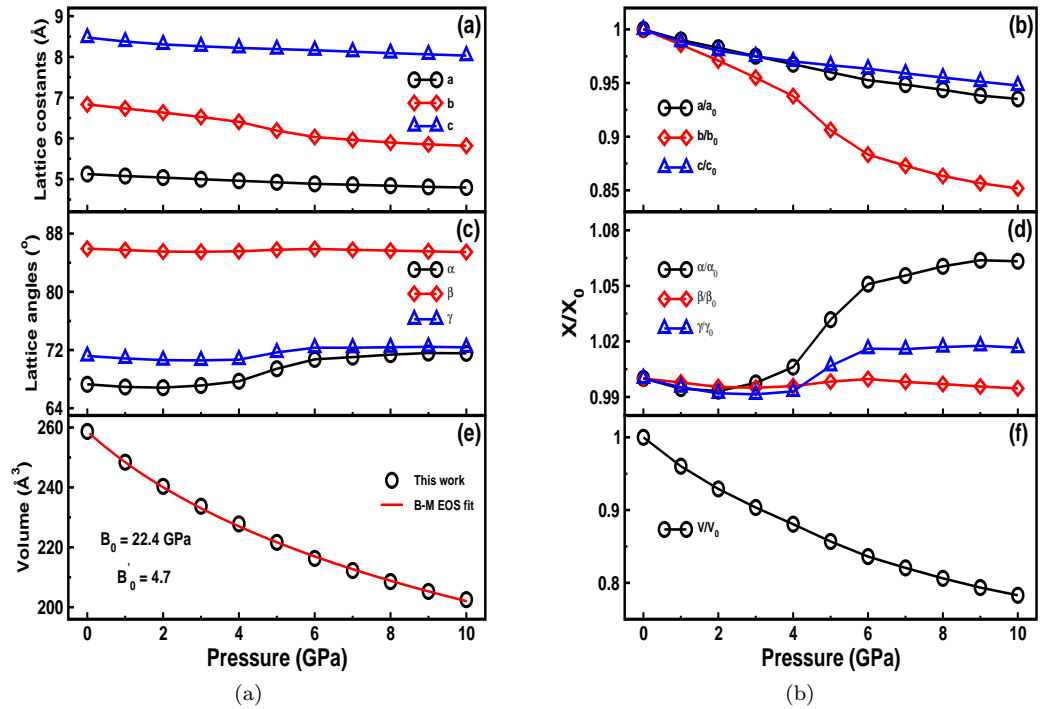

\centering
{\subfigure[]{\includegraphics[height = 3.5in, width=2.5in]{Fig2a.eps}}} \hspace{0.3in}
{\subfigure[]{\includegraphics[height = 3.5in, width=2.5in]{Fig2b.eps}}}
\caption{(Colour online) Calculated (a) lattice constants (a, b, and c), (b) normalized lattice constants (a/a$_0$, b/b$_0$, and c/c$_0$), (c) lattice angles ($\alpha$, $\beta$, and $\gamma$), (d) normalized lattice angles ($\alpha$/$\alpha_0$, $\beta$/$\beta_0$, and $\gamma$/$\gamma_0$) (e) Volume (V), and (f) normalized Volume (V/V$_0$) of K$_2$DNABT as a function of pressure.}
\end{figure*}

\begin{figure*}[h]
\centering
\includegraphics[height = 4.5in, width=4.0in]{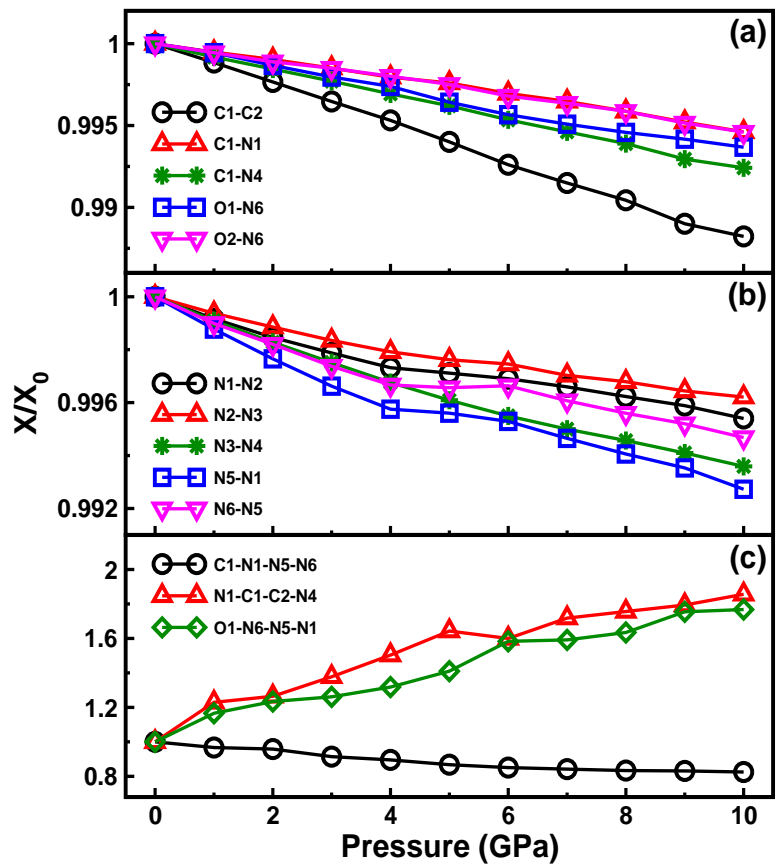}
\caption{(Colour online) Calculated normalized (a, b) bond lengths and (c) torsion angles of K$_2$DNABT as a function of pressure.}
\end{figure*}

\begin{figure*}[h]
\centering
{\subfigure[]{\includegraphics[height = 2.8in, width=1.9in]{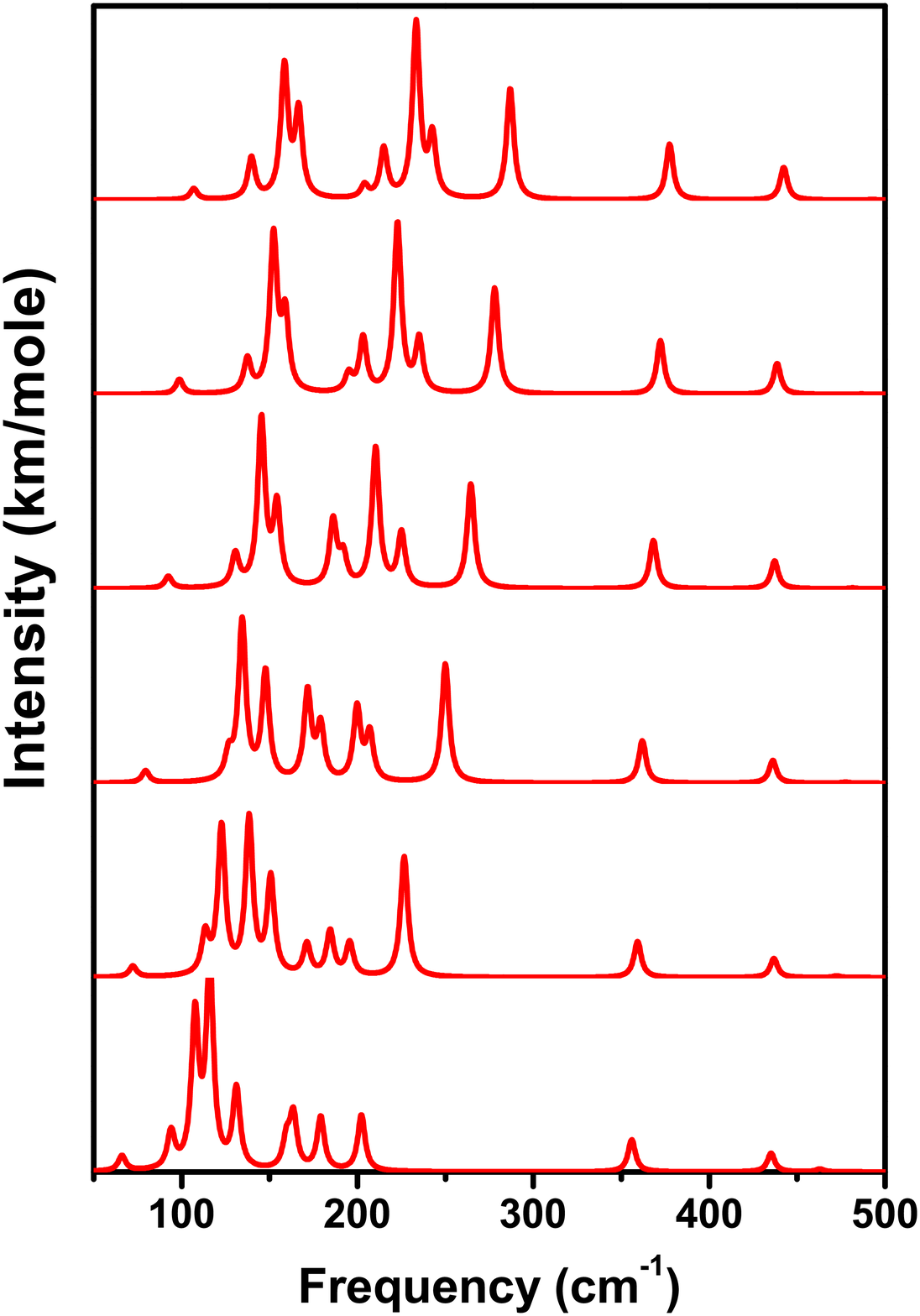}}}
{\subfigure[]{\includegraphics[height = 2.5in, width=1.7in]{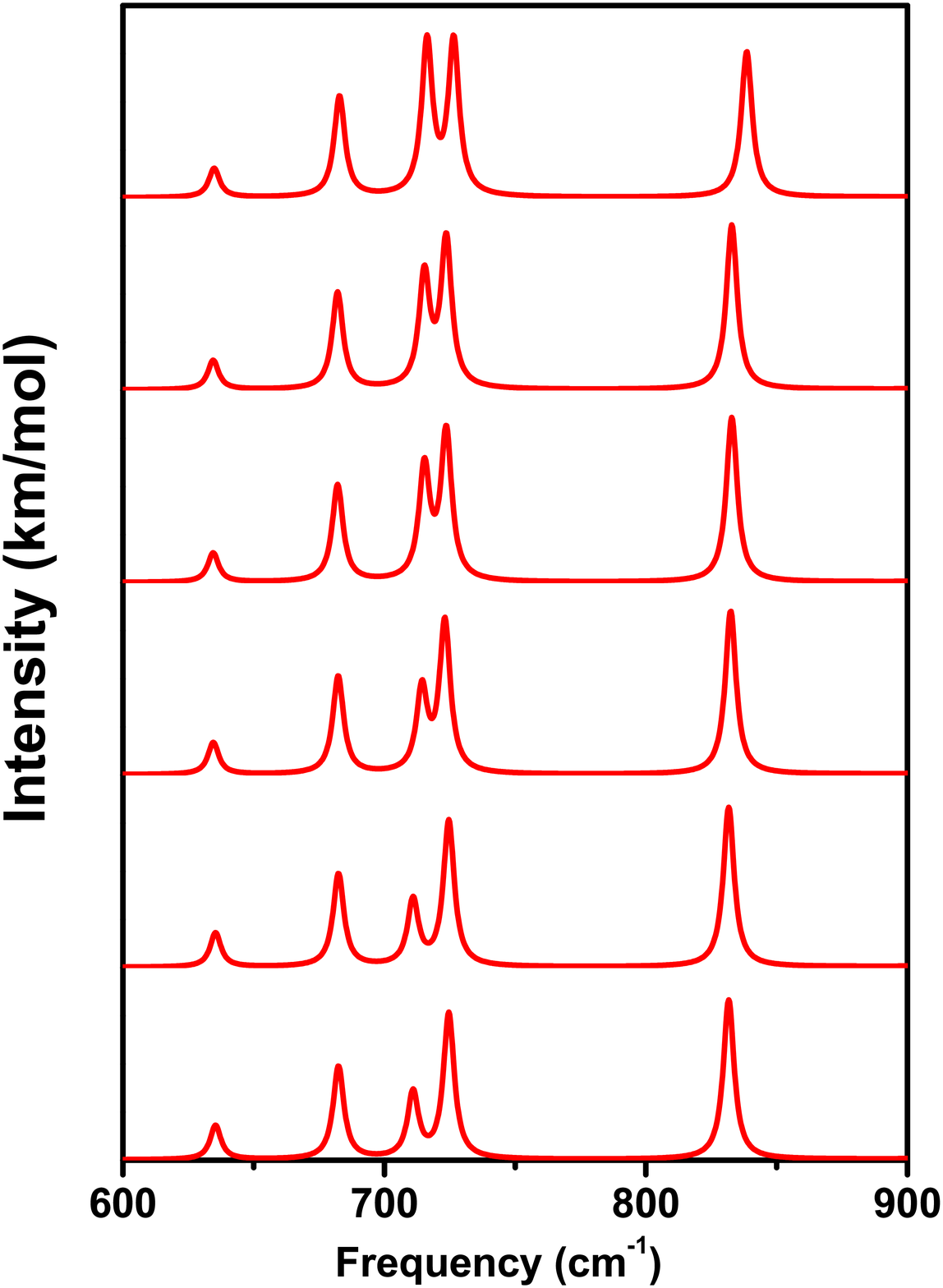}}}
{\subfigure[]{\includegraphics[height = 2.5in, width=1.7in]{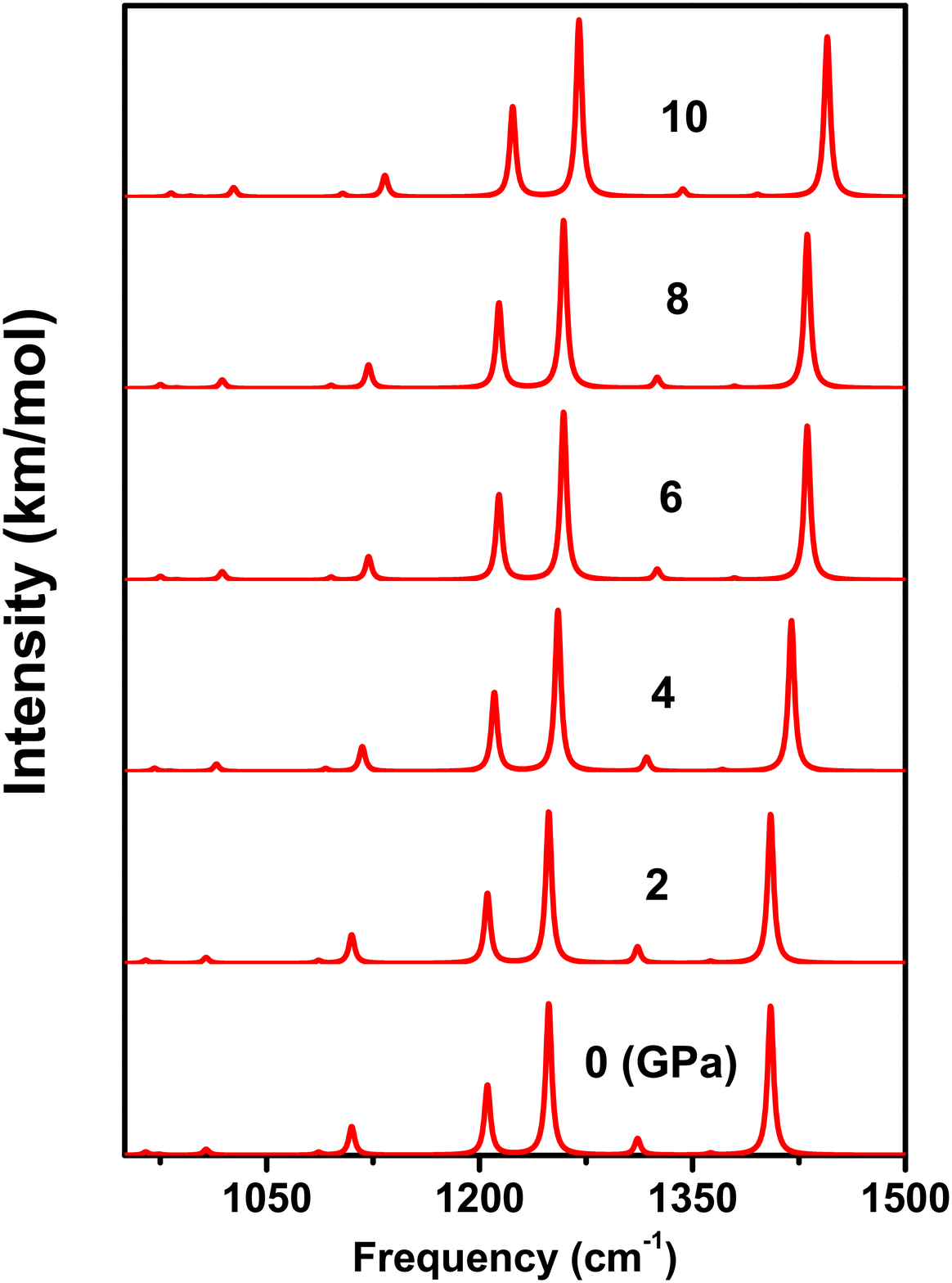}}}
\caption{(Colour online) Calculated IR spectra (a) 40-500 cm$^{-1}$, (b) 600-900 cm$^{-1}$ and (c) 900-1500 cm$^{-1}$ of K$_2$DNABT as a function of pressure.}
\end{figure*}

\begin{figure*}[h]
\centering
{\subfigure[]{\includegraphics[height = 3.4in, width=2.7in]{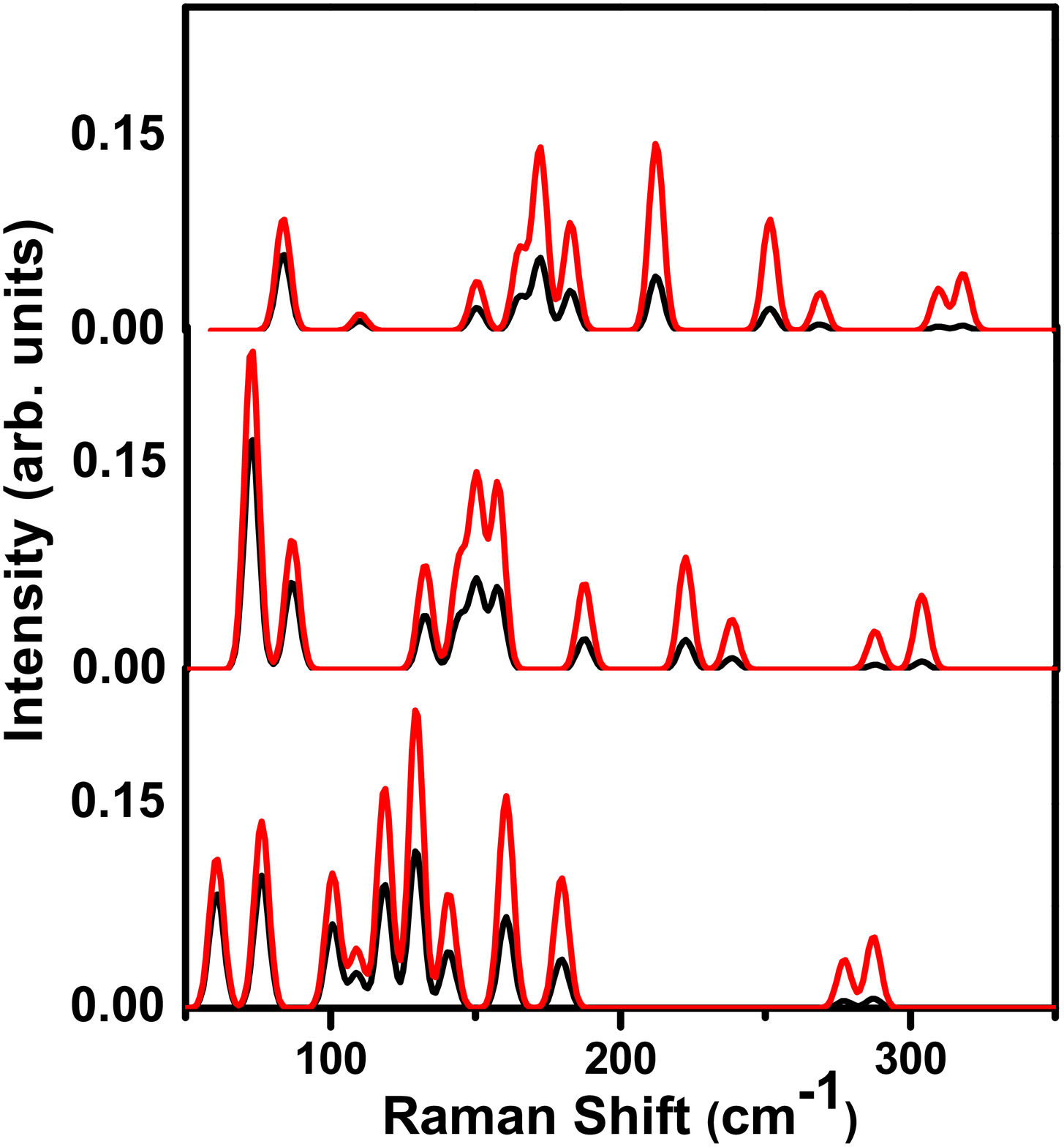}}} \hspace{0.3in}
{\subfigure[]{\includegraphics[height = 3.0in, width=2.5in]{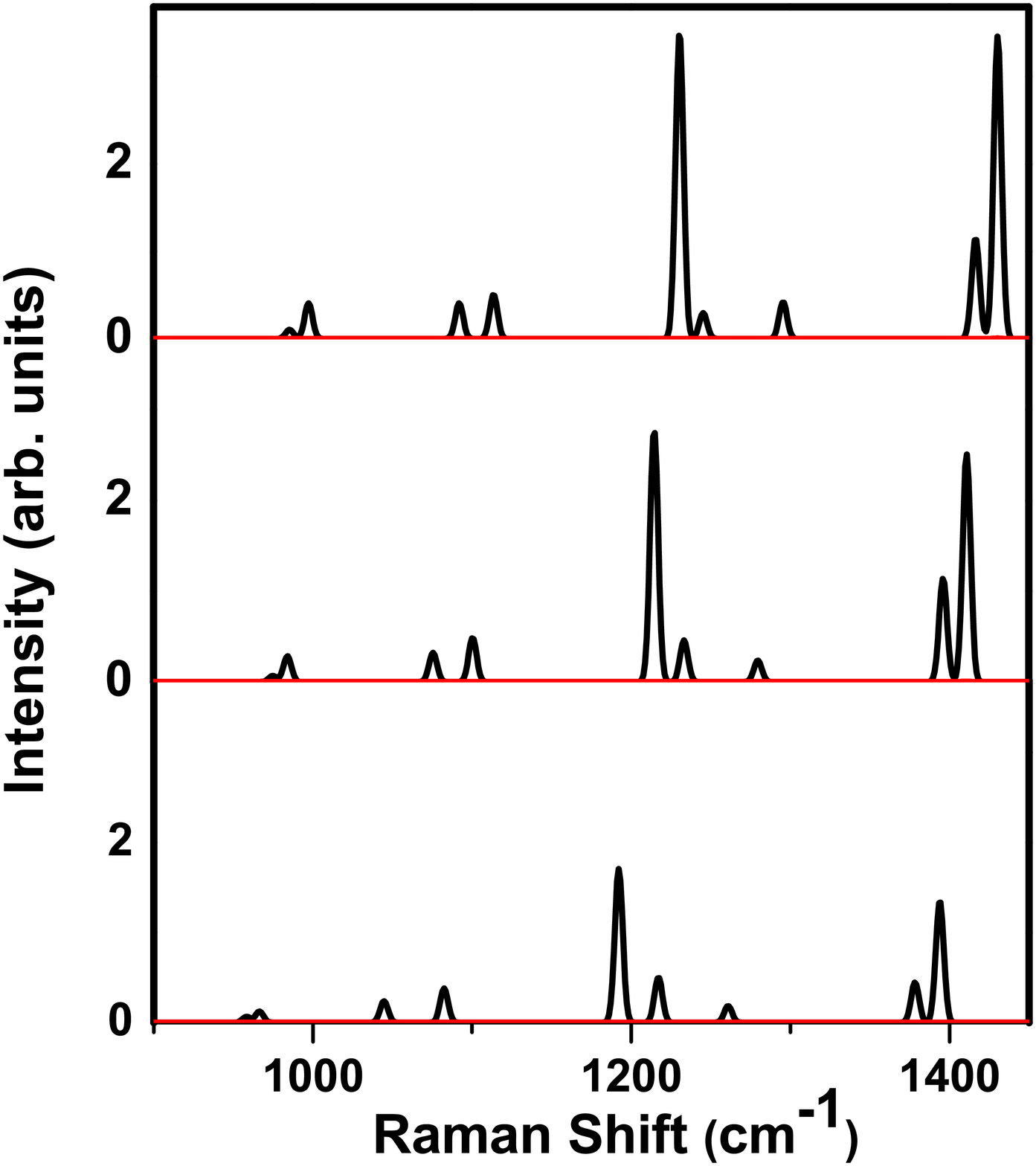}}}
{\subfigure[]{\includegraphics[height = 3.0in, width=2.5in]{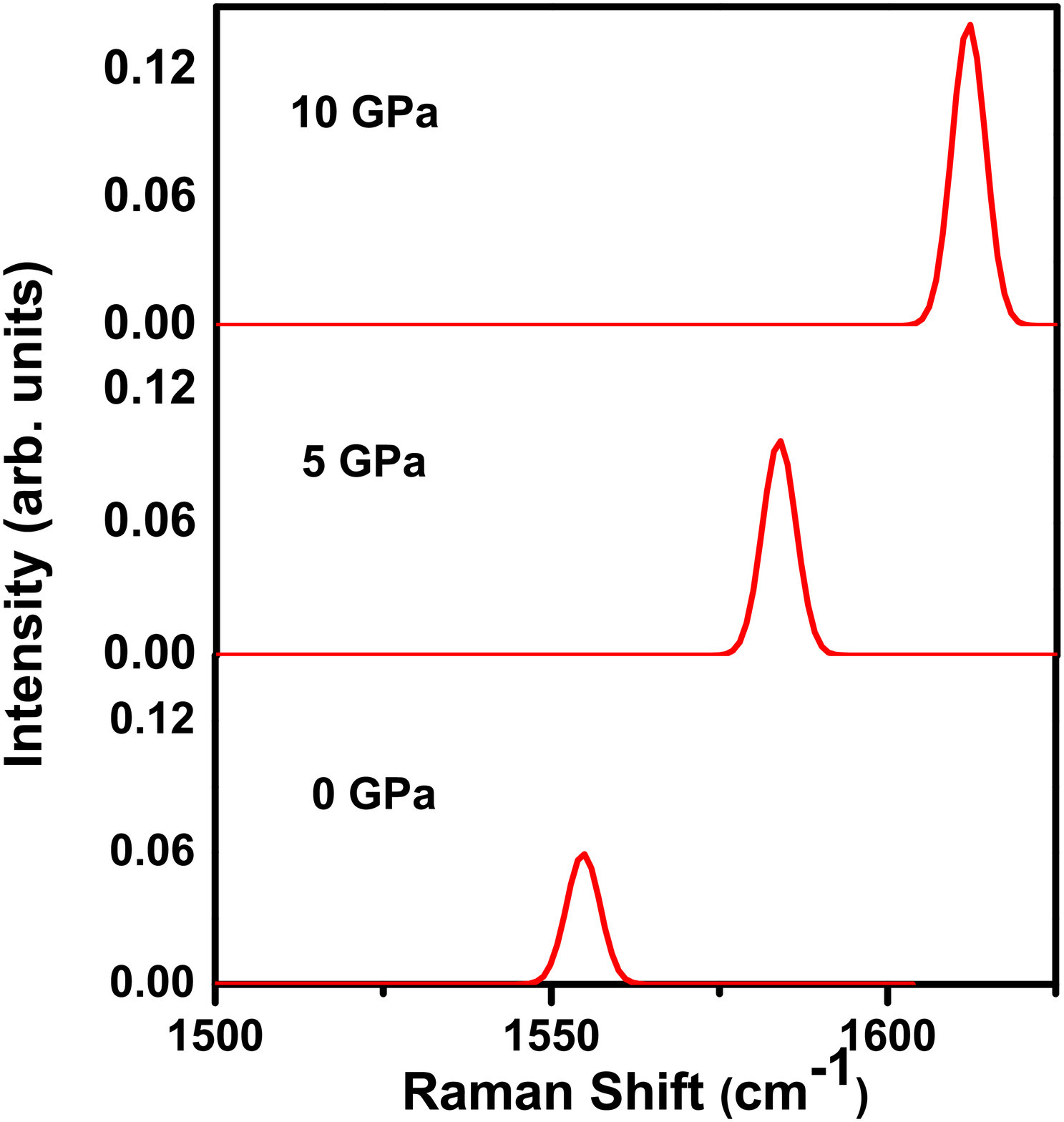}}} \hspace{0.3in}
{\subfigure[]{\includegraphics[height = 3.0in, width=2.5in]{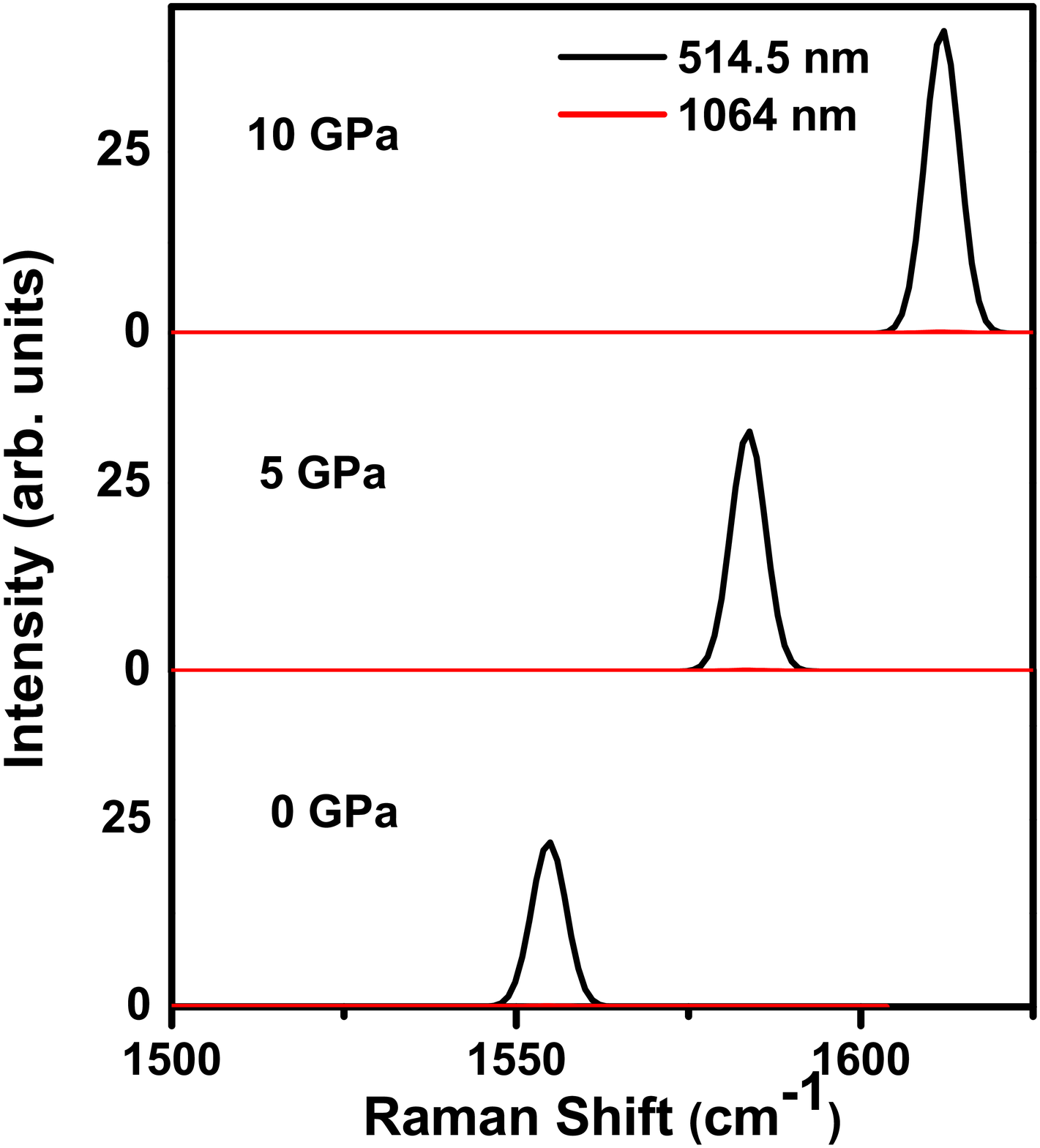}}}
\caption{(Colour online) Calculated Raman spectra (a) 40-350 cm$^{-1}$, (b) 900-1450 cm$^{-1}$ and (c,d) 1500-1625 cm$^{-1}$ of K$_2$DNABT as a function of pressure. Black and red solid lines represent standard Ar (514.5 nm) and Nd:YAG (1064 nm) lasers respectively. The lasers used as an incident light while calculating the Raman spectra of K$_2$DNABT with Gaussian brodening of 2.5 cm$^{-1}$.}
\end{figure*}

\begin{figure*}[h]
\centering
\includegraphics[height = 4.0in, width=5.5in]{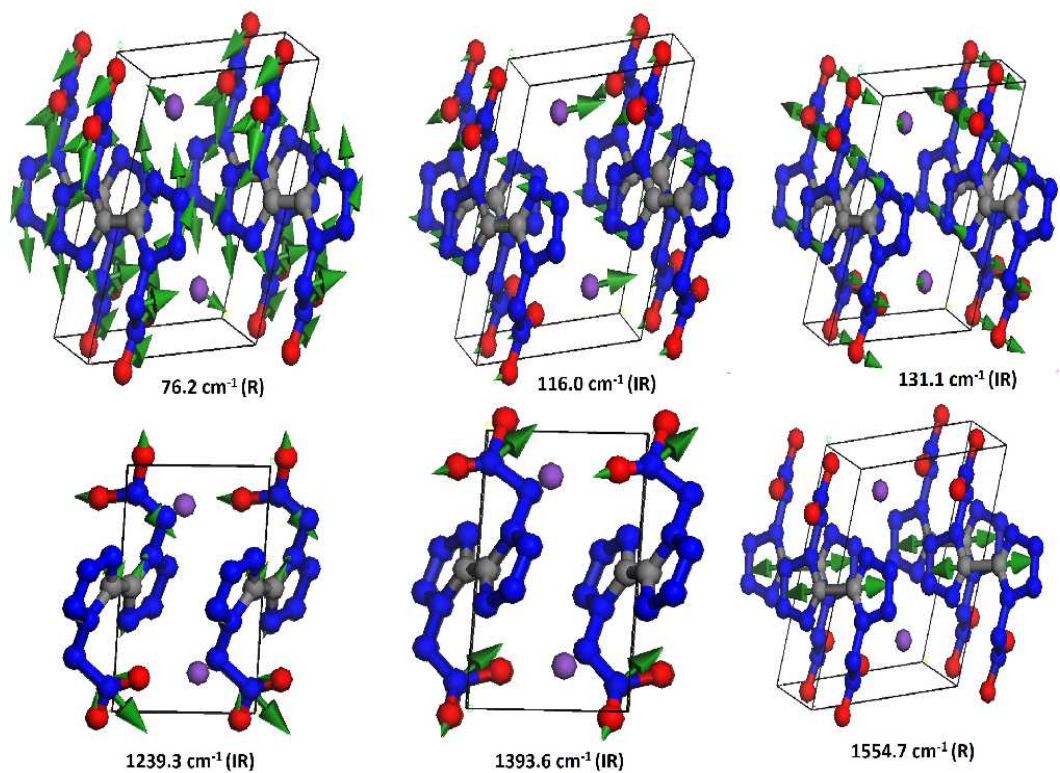}
\caption{(Colour online) The few simulated lattice and internal eigenvectors of  K$_2$DNABT. Here R and IR denote Raman and IR active modes, respectively.}
\end{figure*}

\begin{figure*}[h]
\centering
{\subfigure[]{\includegraphics[height = 3.7in, width=2.5in]{Fig7a.eps}}}
{\subfigure[]{\includegraphics[height = 3.5in, width=2.5in]{Fig7b.eps}}}
{\subfigure[]{\includegraphics[height = 3.5in, width=2.5in]{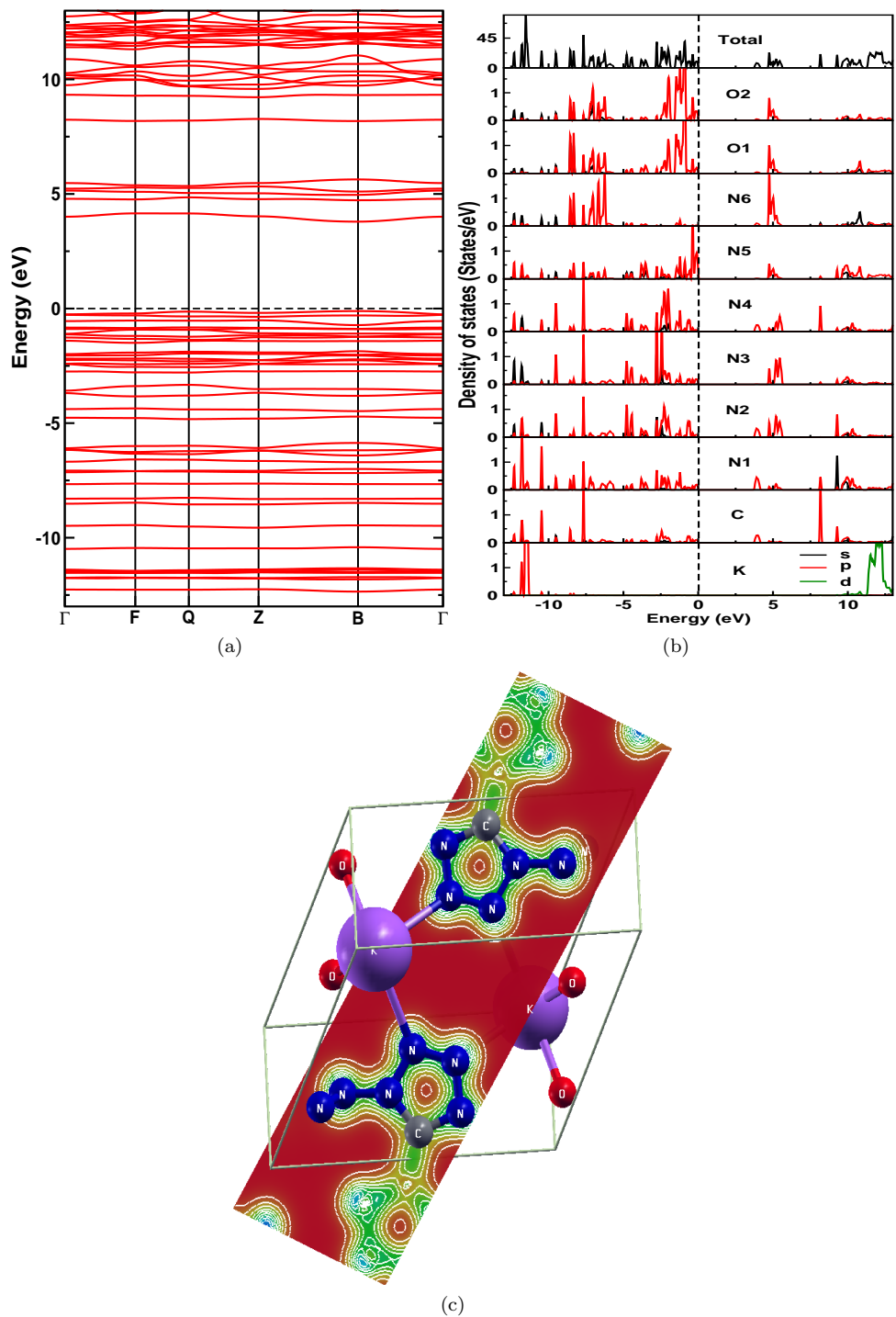}}}
\caption{(Colour online) Calculated (a) Electronic band structure, (b) Partial density of states, (c) Electron charge density of K$_2$DNABT using TB-mBJ potential.}
\end{figure*}

\begin{figure*}[h]
\centering
\includegraphics[height = 3.0in, width=4.5in]{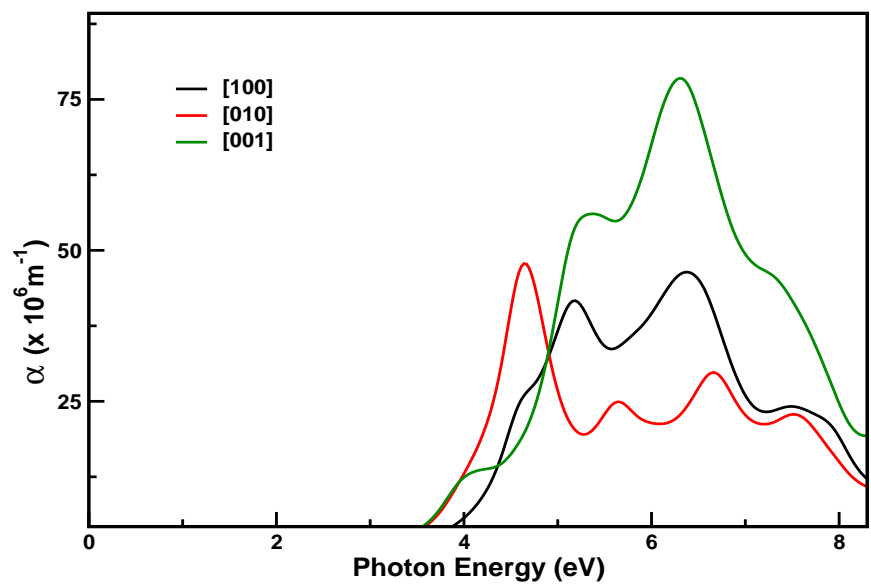}
\caption{(Colour online) Calculated absorption spectra of K$_2$DNABT as function of photon energy using TB-mBJ potential.}
\end{figure*}

\end{document}